# Revisiting the electron-doped SmFeAsO: enhanced superconductivity up to 58.6 K by Th and F codoping

Xiao-Chuan Wang[1,2], Jia Yu[1,2], Bin-Bin Ruan[1,2], Bo-Jin Pan[1,2], Qing-Ge Mu[1,2], Tong Liu[1,2], Kang Zhao[1,2], Gen-Fu Chen[1,2,3], Zhi-An Ren[1,2,3] *

[1] Institute of Physics and Beijing National Laboratory for Condensed Matter Physics, Chinese Academy of Sciences, Beijing 100190, P. R. China

[2] School of Physical Sciences, University of Chinese Academy of Sciences, Beijing 100049, P. R. China

[3] Collaborative Innovation Center of Quantum Matter, Beijing 100190, P. R. China

* Email: renzhian@iphy.ac.cn

In the iron-based high-$T_c$ bulk superconductors, $T_c$ above 50 K was only observed in the electron-doped 1111-type compounds. Here we revisited the electron-doped SmFeAsO polycrystals to make a further investigation for the highest $T_c$ in these materials. To introduce more electron carriers and less crystal lattice distortions, we studied the Th and F codoping effects into the Sm-O layers with heavy electron doping. Dozens of $Sm_{1-x}Th_xFeAsO_{1-y}F_y$ samples were synthesized through the solid state reaction method, and these samples were carefully characterized by the structural, resistive, and magnetic measurements. We found that the codoping of Th and F clearly enhanced the superconducting $T_c$ than the Th or F single-doped samples, with the highest record $T_c$ up to 58.6 K when x = 0.2 and y = 0.225. Further element doping caused more impurities and lattice distortions in the samples with a weakened superconductivity.

Since the discovery of superconductivity in LaFeAsO$_{1-x}$F$_x$ with a superconducting transition temperature ($T_c$) of 26 K, extensive research efforts have been devoted to exploring similar materials due to the relatively high $T_c$, very high values of upper critical field and the similarity to the cuprates.[1–4] By changing the superconducting layers and/or the blocking layers, several new types of iron based superconductors were discovered with the general formulas of 1111-type, 122-type, 111-type, 11-type, *etc.*.[1,5–7] In all these compounds, the 1111-type superconductors have been always holding the record for $T_c$ in bulk materials with several reports from 55 K to 58 K in doped SmFeAsO or GdFeAsO.[3,8–10] Among 122-type superconductors, $T_c$ for hole-doped BaFe$_2$As$_2$ samples is usually below 38 K, while high-$T_c$ above 40 K was observed in surface electron-doped Ba(Fe$_{1-x}$Co$_x$)$_2$As$_2$, and in some rare-earth doped Ca$_{1-x}$RE$_x$Fe$_2$As$_2$ superconductors (RE = rare earth elements), such as 49 K superconductivity in Ca$_{1-x}$Pr$_x$Fe$_2$As$_2$.[5,11,12] In 111-type superconductors, LiFeAs shows superconductivity at 18 K, while $T_c$ of NaFeAs can reach 31 K at 3 GPa.[6,13] The 11-type FeSe becomes superconducting at approximately 8 K while 37 K under high pressure.[7,14] By intercalation of alkali metals into FeSe, superconductivity above 40 K was obtained and a $T_c$ of 48 K was found in K$_{0.8}$Fe$_{1.7}$Se$_2$ at a high pressure.[15,16] Interestingly even higher $T_c$ up to 100 K was also reported for monolayer FeSe thin films.[17] Besides, several other types of iron based superconductors were also found albeit with lower $T_c$, such as Sr$_4$Sc$_2$O$_6$Fe$_2$P$_2$, Sr$_2$VO$_3$FeAs, Ca$_3$Al$_2$O$_{5-y}$Fe$_2$Pn$_2$ (Pn = As and P), Ca$_{10}$(Pt$_3$As$_8$)(Fe$_2$As$_2$)$_5$, Ca$_{1-x}$La$_x$FeAs$_2$, LiFeO$_2$Fe$_2$Se$_2$, CaAFe$_4$As$_4$ (A = K, Rb and Cs), LaFeSiH, *etc.*.[18–25]

The parent compounds of these iron pnictides are usually bad metals with structural and antiferromagnetic transitions at low temperature, and chemical doping or external pressure are the common ways to induce superconductivity in them. As the only type with high $T_c$ above 50 K, the 1111-type bulk superconductors have been widely studied to achieve a record $T_c$ by various doping methods, external pressure or synthesizing techniques.[8,26,27] After the first reported LaFeAsO$_{1-x}$F$_x$ with $T_c$ around 26 K, enhancement in $T_c$ up to 55 K was soon observed in SmFeAsO$_{1-x}$F$_x$ and

SmFeAsO$_{1-\delta}$ by a high-pressure synthesis method.[3,8] Same as F-doping in LaFeAsO$_{1-x}$F$_x$, substitution of Th at rare-earth site and even H doping at the O site also resulted in electron doping with similar $T_c$, while chemical doping in the Fe-As layers always lead to a lowering of $T_c$.[27–30] External physical pressure effects on $T_c$ were also studied in 1111-type superconductors. Pressure enhanced $T_c$ for LaFeAsO$_{0.89}$F$_{0.11}$ (from 26 K to 43 K), while it suppressed $T_c$ for CeFeAsO$_{0.88}$F$_{0.12}$ and REFeAsO$_{0.85}$ (RE = Sm and Nd).[26,31,32] Moreover, superconductivity at 57.8 K was found in SmFeAsO$_{1-x}$F$_x$ film grown by molecular beam epitaxy.[33] Recently, several groups found that low temperature sintering and slow cooling techniques could introduce high F doping levels with $T_c$ enhanced up to 58 K in SmFeAsO$_{1-x}$F$_x$.[10,34,35]

Considering the elements doping by Th or F both can introduce electron carriers into the Sm-O layers while keeping intact for the Fe-As layers, here we revisited the SmFeAsO by simultaneous codoping of Th and F elements, with the intention to further increase electron doping levels while maintaining the crystal structure with less distortions. We synthesized dozens of the Th and F codoped Sm$_{1-x}$Th$_x$FeAsO$_{1-y}$F$_y$ samples by the solid state reaction and carefully characterized their superconducting properties. We found that the codoping of Th and F can effectively enhance the superconductivity and lead to a maximum $T_c$ of 58.6 K.

The Th and F codoped polycrystalline Sm$_{1-x}$Th$_x$FeAsO$_{1-y}$F$_y$ samples were prepared by a conventional solid state reaction method. At first, SmAs and FeAs were pre-sintered from Sm, Fe and As powders sealed in evacuated quartz tubes at 900 °C for 30 h. These compounds were ground into powders for later use. Then the fine powders of SmAs, FeAs, Fe$_2$O$_3$, ThO$_2$, FeF$_2$ and Fe were mixed together according to the nominal stoichiometric ratio of Sm$_{1-x}$Th$_x$FeAsO$_{1-y}$F$_y$, ground thoroughly in an agate mortar and pressed into small pellets. All the preparation processes were carried out in an argon protected glove box. The pellets were sealed into evacuated quartz tubes and heated at 1150 °C for 30 h, then slowly cooled down to 600 °C in 100 h. Finally the quartz tubes were cooled in the furnace after shutting off the power.

The samples were characterized by powder X-ray diffraction (XRD) method on a PAN-analytical X-ray diffractometer with Cu-K$_\alpha$ radiation at room temperature for crystal structure determination and chemical phase analysis. The temperature dependence of resistivity was measured by the standard four-probe method using a PPMS system (Physical Property Measurement System, Quantum Design). The temperature dependence of DC magnetic susceptibility was measured using an MPMS system (Magnetic Property Measurement System, Quantum Design).

To examine the Th and F codoping effects in the 1111 phase, all the $Sm_{1-x}Th_xFeAsO_{1-y}F_y$ polycrystalline samples were characterized by the room temperature XRD analysis, and some of the typical diffraction patterns are shown in Fig. 1. The expanded view near the (102) diffraction peak is depicted in the right panel. The undoped parent compound SmFeAsO shows a clean tetragonal ZrCuSiAs-type structure with no impurity phase, and the F-doped $SmFeAsO_{0.85}F_{0.15}$ shows nearly pure phase with tiny SmOF observed, while considerable impurities of $ThO_2$ and SmAs are noticeable in the Th-doped $Sm_{0.85}Th_{0.15}FeAsO$ due to the difficulty of Th-doping. These XRD patterns and the appearance of impurities are similar with previous reports.[2,27,28] For the Th and F codoped $Sm_{1-x}Th_xFeAsO_{1-y}F_y$ samples, the main tetragonal 1111 phase can be clearly characterized in all samples in spite of the growing of the impurity phases of $ThO_2$, SmOF, SmAs, and $Fe_2As$, which become dominant and surpass the 1111 phase when x > 0.2 and y > 0.2. This also indicates that Th or F cannot be fully doped at the Sm site or O site, and the real electron doping level should be smaller than the nominal one. The shift of the (102) diffraction peak to the higher angle with increasing Th and F doping indicates the decrease of the lattice constants, and confirms the electron doping.

In Fig. 2 we show the temperature dependence of resistivity for some typical doped $Sm_{1-x}Th_xFeAsO_{1-y}F_y$ samples within the temperature range from 2 K to 300 K, and the expanded view near $T_c$ is shown in the right panel. As the original SDW transition in the undoped SmFeAsO was suppressed by doping, all the resistivity curves show metallic behaviors above the superconducting transitions for the doped samples. In our experiments, the ambient-pressure synthesized single-doped $SmFeAsO_{0.85}F_{0.15}$ has an onset $T_c$ of 52.8 K, which is slightly lower than the 55 K $T_c$ of our previously reported high-pressure synthesized $SmFeAsO_{0.9}F_{0.1}$.[3] The Th-doped $Sm_{0.85}Th_{0.15}FeAsO$ has a $T_c$ of 45.0 K, and this relatively lower $T_c$ is mainly due to the less electron doping concentration from the difficulty of Th-doping, which is exposed by the remained $ThO_2$ impurity phase in the sample. For the Th and F codoped $Sm_{1-x}Th_xFeAsO_{1-y}F_y$ samples, the nominal electron doping level is defined as n = x + y, since the Th and F dopants each introduce one electron per atom. Here we mainly studied the heavily electron doped samples. Clearly, the superconducting transition occurs at 55.2 K for $Sm_{0.9}Th_{0.1}FeAsO_{0.9}F_{0.1}$, slightly higher than the $T_c$ value of $Sm_{0.85}Th_{0.15}FeAsO$ or $SmFeAsO_{0.85}F_{0.15}$, and these results are similar to the previous report.[27,28] For all these Th and F codoped SmFeAsO samples with n > 0.2, the onset $T_c$ surpasses 55 K and gradually increases with a maximum value of $T_c$ ~ 58.6 K for the sample with a nominal composition of $Sm_{0.8}Th_{0.2}FeAsO_{0.775}F_{0.225}$, then drops slowly with further electron doping as impurity phase increases. The enhancement of superconductivity can be attributed to the introducing of more electron carriers and less crystal lattice distortions by the codoping comparing with other doping methods. The $T_c$ around 58.6 K was reproducibly observed in several samples with the electron doping level n ~ 0.425, which reveals the highest $T_c$ in the 1111-type SmFeAsO by Th and F codoping.

The DC magnetization data for all the $Sm_{1-x}Th_xFeAsO_{1-y}F_y$ samples were measured between 2 K and 70 K with both zero-field-cooling (ZFC) and field-cooling (FC) modes under a magnetic field of 10 Oe, as shown in Fig. 3 for several typical samples, with the right panel shows the expanded view close to the magnetic transition. All the samples show onset magnetic $T_c$ above 50 K except the Th single-doped $Sm_{0.85}Th_{0.15}FeAsO$, and all the Th and F codoped samples have higher $T_c$ than the F single-doped sample $SmFeAsO_{0.85}F_{0.15}$. The onset $T_c$ gradually increases with increasing electron doping level n, then saturates for further increasing n. The highest $T_c$ from the magnetic transitions was observed in the

$Sm_{0.8}Th_{0.2}FeAsO_{0.775}F_{0.225}$ sample, with an onset $T_c$ ~ 58.0 K, and similar $T_c$ was detected in several samples. The diamagnetic superconducting transitions are consistent with the resistivity superconducting transitions, and show sharp transitions and bulk behavior with high superconducting volume fraction for the codoped samples. The decrease of the superconducting volume fraction in high doping level samples is mainly owing to the increase of remained impurity phases.

To further elucidate the superconducting properties, the relationships for the lattice parameters $a$ and $c$, the superconducting $T_c$ versus the nominal electron doping level n for all the $Sm_{1-x}Th_xFeAsO_{1-y}F_y$ samples were summarized respectively, as plotted in Fig. 4. Along doping when n < 0.3, the crystal structure has a quick shrinkage along both $a$-axis and $c$-axis, and the $T_c$ also rises rapidly with the increase of doping level. The codoping effects are similar to the previous report of low-level doping,[27] while with a higher $T_c$. When n > 0.3, the lattice parameters shrink little along $a$-axis while more along $c$-axis, and accordingly, the resistive $T_c$ becomes almost saturated around 58 K. The magnetic $T_c$ shows similar behavior with the resistive one by a little lower transition temperature. The optimal doping level is at n ~ 0.425 with a maximum $T_c$ ~ 58.6 K. For more elements doping of Th and F, the impurity phases become dominant and lattice distortions in the 1111 phase may turn into a main obstruction for the enhancement of superconductivity, and even causes the destruction of superconductivity with no zero resistivity for n > 0.5.

In summary, we synthesized the Th and F codoped $Sm_{1-x}Th_xFeAsO_{1-y}F_y$ samples by the solid state reaction method. By systematically investigating the evolution of crystal structure and superconducting properties, we found the obvious enhancement of superconductivity by the codoping effect than the Th or F single-doped samples, which lead to a highest superconducting $T_c$ of 58.6 K in the sample $Sm_{0.8}Th_{0.2}FeAsO_{0.775}F_{0.225}$.

The authors are grateful for the financial supports from the National Natural Science Foundation of China (No. 11474339), the National Basic Research Program

[1] Kamihara Y, Watanabe T, Hirano M and Hosono H 2008 *J. Am. Chem. Soc.* **130** 3296
[2] Chen X H, Wu T, Wu G, Liu R H, Chen H and Fang D F 2008 *Nature* **453** 761
[3] Ren Z A, Lu W, Yang J, Yi W, Shen X L, Li Z C, Che G C, Dong X L, Sun L L, Zhou F and Zhao Z X 2008 *Chin. Phys. Lett.* **25** 2215
[4] Jaroszynski J, Riggs S C, Hunte F, Gurevich A, Larbalestier D C and Boebinger G S 2008 *Phys. Rev.* B **78** 064511
[5] Rotter M, Tegel M and Johrendt D 2008 *Phys. Rev. Lett.* **101** 107006
[6] Wang X C, Liu Q Q, Lv Y X, Gao W B, Yang L X, Yu R C, Li F Y and Jin C Q 2008 *Solid State Commun.* **148** 538
[7] Hsu F C, Luo J Y, Yeh K W, Chen T K, Huang T W, Wu P M, Lee Y C, Huang Y L, Chu Y Y, Yan D C and Wu M K 2008 *Proc. Natl. Acad. Sci. USA* **105** 14262
[8] Ren Z A, Che G C, Dong X L, Yang J, Lu W, Yi W, Shen X L, Li Z C, Sun L L, Zhou F and Zhao Z X 2008 *Europhys. Lett.* **83** 17002
[9] Wang C, Li L J, Chi S, Zhu Z W, Ren Z, Li Y K, Wang Y T, Lin X, Luo Y K, Jiang S, Xu X F, Cao G H and Xu Z A 2008 *Europhys. Lett.* **83** 67006
[10] Fujioka M, Denholme S J, Ozaki T, Okazaki H, Deguchi K, Demura S, Hara H, Watanabe T, Takeya H, Yamaguchi T, Kumakura H and Takano Y 2013 *Supercond. Sci. Technol.* **26** 085023
[11] Kyung W S, Huh S S, Koh Y Y, Choi K -Y, Nakajima M, Eisaki H, Denlinger J D, Mo S -K, Kim C and Kim Y K 2016 *Nat. Mater.* **15** 1233
[12] Lv B, Deng L, Gooch M, Wei F, Sun Y, Meen J K, Xue Y Y, Lorenz B and Chu C W 2011 *Proc. Natl. Acad. Sci. USA* **108** 15705
[13] Zhang S J, Wang X C, Liu Q Q, Lv Y X, Yu X H, Lin Z J, Zhao Y S, Wang L, Ding Y, Mao H K and Jin C Q 2009 *Europhys. Lett.* **88** 47008
[14] Medvedev S, Mcqueen T M, Troyan I A, Palasyuk T, Eremets M I, Cava R J, Naghavi S, Casper F, Ksenofontov V, Wortmann G and Felser C 2009 *Nat. Mater.* **8** 630
[15] Guo J G, Jin S F, Wang G, Wang S C, Zhu K X, Zhou T T, He M and Chen X L 2010 *Phys. Rev.* B **82** 180520
[16] Sun L L, Chen X J, Guo J, Gao P W, Huang Q Z, Wang H D, Fang M H, Chen X L, Chen G F, Wu Q, Zhang C, Gu D C, Dong X L, Wang L, Yang K, Li A G, Dai X, Mao H -K and Zhao Z X 2012 *Nature* **483** 67
[17] Ge J F, Liu Z L, Liu C H, Gao C L, Qian D, Xue Q K, Liu Y and Jia J F 2015 *Nat. Mater.* **14** 285
[18] Ogino H, Matsumura Y, Katsura Y, Ushiyama K, Horii S, Kishio K and Shimoyama J 2009 *Supercond. Sci. Technol.* **22** 075008
[19] Zhu X Y, Han F, Mu G, Cheng P, Shen B, Zeng B and Wen H H 2009 *Phys. Rev.* B **79** 220512
[20] Shirage P M, Kihou K, Lee C H, Kito H, Eisaki H and Iyo A 2011 *J. Am. Chem. Soc.* **133** 9630
[21] Kakiya S, Kudo K, Nishikubo Y, Oku K, Nishibori E, Sawa H, Yamamoto T, Nozaka T and Nohara M 2011 *J. Phys. Soc. Jpn.* **80** 093704
[22] Katayama N, Kudo K, Onari S, Mizukami T, Sugawara K, Sugiyama Y,

Kitahama Y, Iba K, Fujimura K, Nishimoto N, Nohara M and Sawa H 2013 *J. Phys. Soc. Jpn.* **82** 123702
[23] Lu X F, Wang N Z, Zhang G H, Luo X G, Ma Z M, Lei B, Huang F Q and Chen X H 2014 *Phys. Rev.* B **89** 020507
[24] Iyo A, Kawashima K, Kinjo T, Nishio T, Ishida S, Fujihisa H, Gotoh Y, Kihou K, Eisaki H and Yoshida Y 2016 *J. Am. Chem. Soc.* **138** 3410
[25] Bernardini F, Garbarino G, Sulpice A, Regueiro M N, Gaudin E, Chevalier B, Cano A and Tencé S 2017 arXiv:1701.05010 [cond-mat.supr-com]
[26] Takahashi H, Igawa K, Arii K, Kamihara Y, Hirano M and Hosono H 2008 *Nature* **453** 376
[27] Li Y K, Lin X, Tao Q, Chen H, Wang C, Li L J, Luo Y K, He M, Zhu Z W, Cao G H and Xu Z A 2009 *Chin. Phys. Lett.* **26** 017402
[28] Zhigadlo N D, Katrych S, Weyeneth S, Puzniak R, Moll P J W, Bukowski Z, Karpinski J, Keller H and Batlogg B 2010 *Phys. Rev.* B **82** 064517
[29] Hanna T, Muraba Y, Matsuishi S, Igawa N, Kodama K, Shamoto S and Hosono H 2011 *Phys. Rev.* B **84** 024521
[30] Sefat A S, Huq A, McGuire M A, Jin R Y, Sales B C, Mandrus D, Cranswick L M D, Stephens P W and Stone K H 2008 *Phys. Rev.* B **78** 104505
[31] Zocco D A, Hamlin J J, Baumbach R E, Maple M B, Mcguire M A, Sefat A S, Sales B C, Jin R, Mandrus D, Jeffries J R, Weir S T and Vohra Y K 2008 *Physica* C **468** 2229
[32] Yi W, Sun L L, Ren Z A, Lu W, Dong X L, Zhang H J, Dai X, Fang Z, Li Z C, Che G C, Yang J, Shen X L, Zhou F and Zhao Z X 2008 *Europhys. Lett.* **83** 57002
[33] Ueda S, Takeda S, Takano S, Yamamoto A and Naito M 2011 *Appl. Phys. Lett.* **99** 232505
[34] Wang C L, Gao Z S, Wang L, Qi Y P, Wang D L, Yao C, Zhang Z Y and Ma Y W 2010 *Supercond. Sci. Technol.* **23** 055002
[35] Singh S J, Shimoyama J, Yamamoto A, Ogino H and Kishio K 2013 *IEEE Trans. Appl. Supercond.* **23** 7300605

**Figure Captions:**

Figure 1: Powder XRD patterns at room temperature for several typical $Sm_{1-x}Th_xFeAsO_{1-y}F_y$ samples, with an expanded view near the (102) peaks placed in the right panel.

Figure 2: The temperature dependence of resistivity for some typical doped $Sm_{1-x}Th_xFeAsO_{1-y}F_y$ samples. The right panel shows the enlarged view around the superconducting transition.

Figure 3: The temperature dependence of magnetic susceptibility for several typical $Sm_{1-x}Th_xFeAsO_{1-y}F_y$ samples between 2 K and 70 K with both ZFC and FC measurements. The right panel shows the expanded curve at the superconducting transition.

Figure 4: The relationship between the nominal electron doping level n and (a) the lattice parameters *a* and *c*, (b) onset $T_c$ determined by resistive and magnetic measurements for all the $Sm_{1-x}Th_xFeAsO_{1-y}F_y$ samples.

Fig. 1.

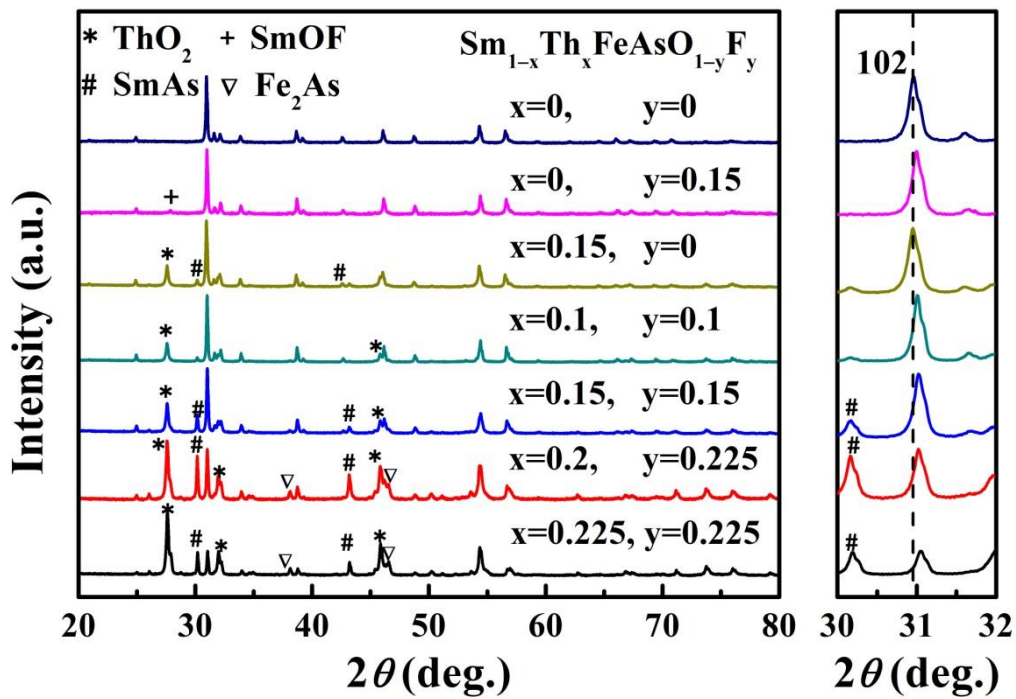

**Fig. 2.**

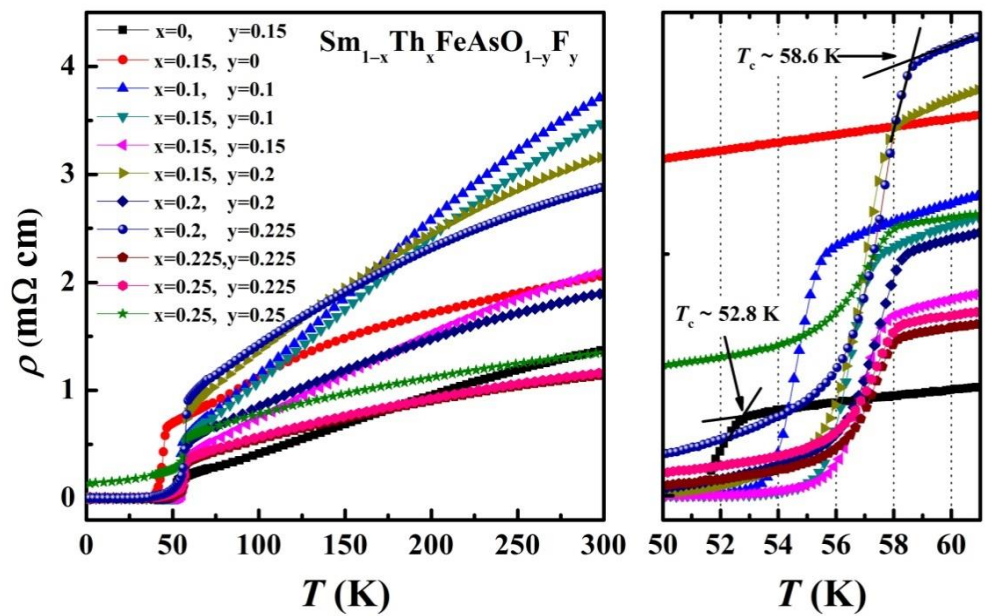

**Fig. 3.**

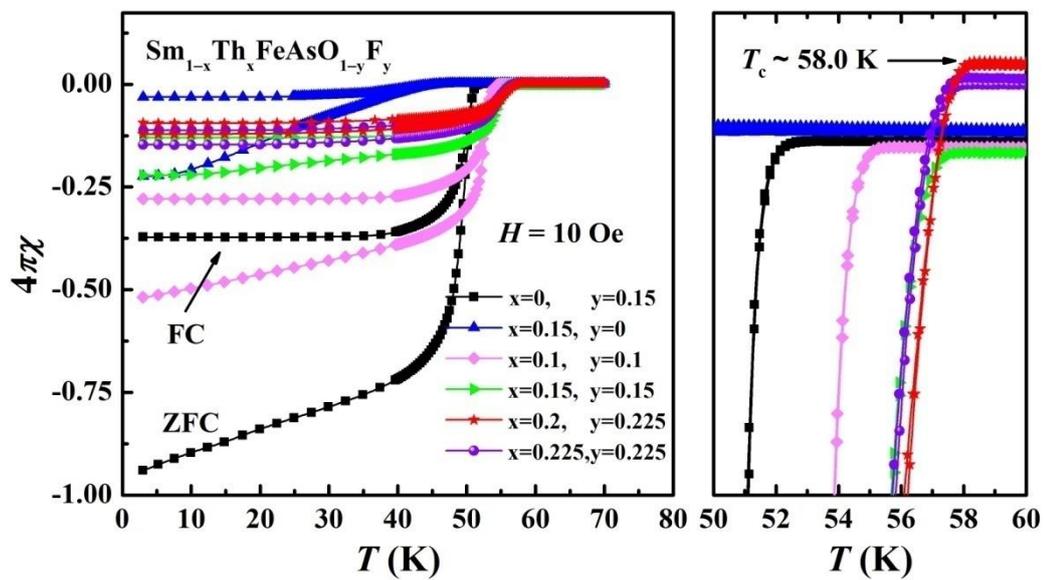

**Fig. 4.**

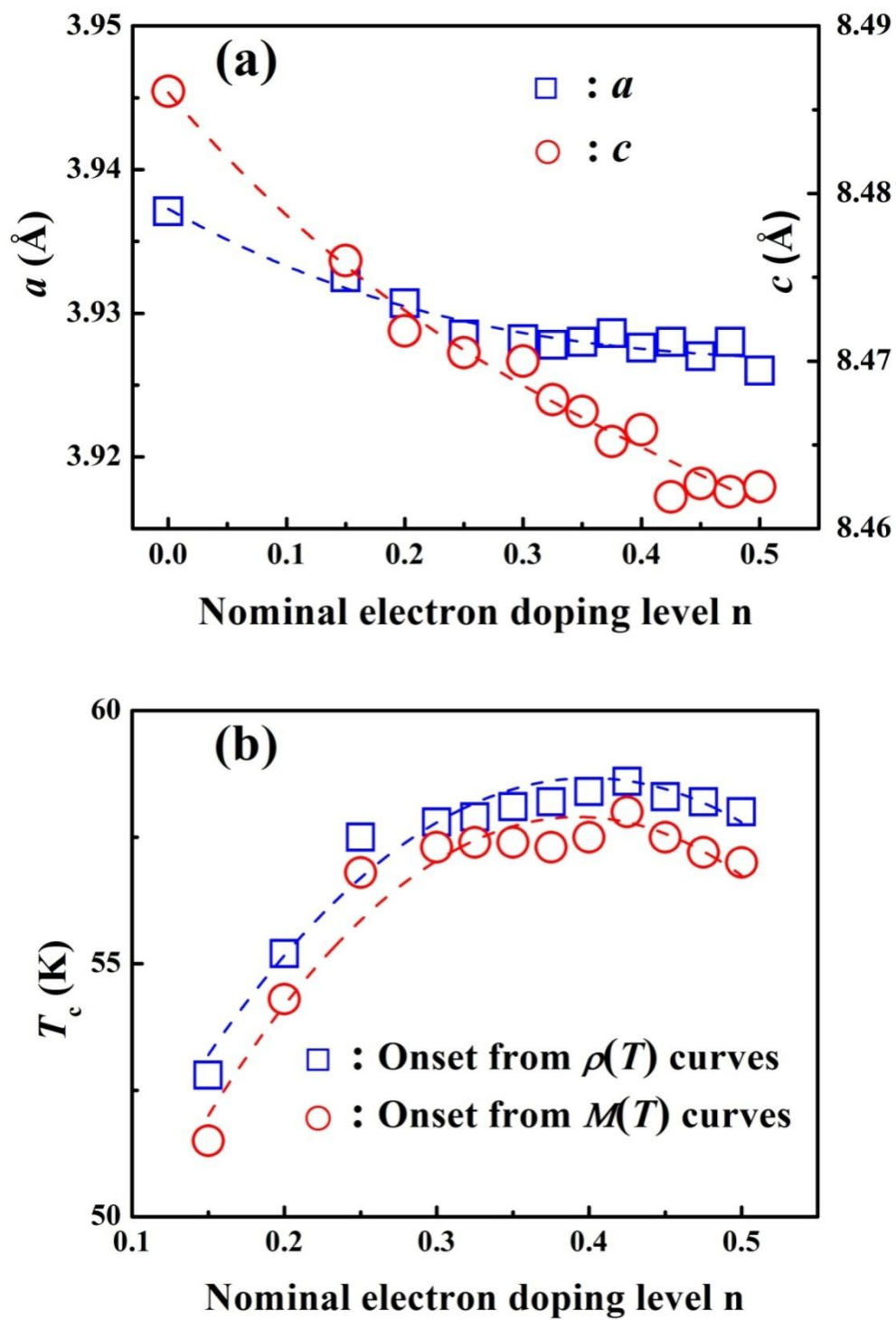